%Paper: astro-ph/9404038
%From: "Peter Meszaros" <nnp@astro.psu.edu>
%Date: Mon, 18 Apr 94 09:51:53 EDT

% wind3.tex,
\def\etal{{\it et~al.}}

\def\halfspace{\baselineskip=12pt plus .1pt}

\def\papersize{\magnification=1200}  %all fonts are normal size.

\font\tfont=cmmib10
\newfam\vecfam

\textfont\vecfam=\tfont \scriptfont\vecfam=\seveni
\scriptscriptfont\vecfam=\fivei

\parindent=40pt
\settabs 7 \columns
\tolerance=1600
\parskip 1ex
% PAGE NUMBERING : DOESN'T NUMBER THE FIRST PAGE
\def\foolit{\ifnum\pageno > 1 \number\pageno\fi}
%\raggedright
\raggedbottom
\def\frac#1#2{{#1\over#2}}
\def\Mesz{M\'esz\'aros\ }
\def\Pacz{Paczy\'nski\ }

\def\ctl{\centerline}

\def\lbr{ \hfill\break }

\def\ref{\par \noindent \hangindent=2pc \hangafter=1 }
\def\etal{{\it et~al.\ }}
\def\mathnew{\mathsurround=0pt}
\def\simov#1#2{\lower .5pt\vbox{\baselineskip0pt \lineskip-.5pt
	\ialign{$\mathnew#1\hfil##\hfil$\crcr#2\crcr\sim\crcr}}}
\def\simg{\mathrel{\mathpalette\simov >}}
\def\siml{\mathrel{\mathpalette\simov <}}
\def\lambdabar{\mathrel{\lower 1pt\hbox{$\mathchar'26$}\mkern-9mu
        \hbox{$\lambda$}}}

\def\msun{{\,M_\odot}}

\def\cm{{\rm\,cm}}

\def\msk{\vskip 2ex\noindent}
\def\bsk{\vskip 3ex\noindent}
\def\msun{M_{\odot}}

%
% UNITS:

\def\cm{~\rm{cm}}

\def\s{~\rm{s}}
\def\si{~ {\rm s}^{-1} }

\def\erg{~\rm{ergs}}
\def\ergs{~\rm{ergs}}

\def\G{~\rm{G}}

\papersize
\halfspace
%\doublespace
%\preprintsize
%
$~$\bsk
%\vskip 0.5cm
\ctl{\bf UNSTEADY OUTFLOW MODELS FOR COSMOLOGICAL GAMMA-RAY BURSTS}
\bsk
\ctl{ M.J. Rees$^{1}$ and P. \Mesz$^{2}$}
\bsk
\ctl{$^1$~ Institute of Astronomy, Madingley Road, Cambridge CB3 OHA, England}
\ctl{$^2$~ Pennsylvania State University, 525 Davey Lab, University Park, PA
16803}
\bsk
\ctl{ Submitted to Ap.J.(Letters): ~4/5/94~~~;~ accepted:~~~~~ }
\bsk
\ctl{\bf Abstract}
\bsk
The 'event' that triggers  a gamma ray burst cannot last for more than a few
seconds. This is, however, long compared with  the dynamical timescale of a
compact
stellar-mass object ($\sim 10^{-3}$ seconds). Energy is assumed to be released
as an outflow with high mean lorentz factor $\Gamma$. But a compact
stellar-mass
collapse or merger is, realistically, likely to generate a  mass (or energy)
flux
that is unsteady  on  some timescales in the range $10^{-3}$ - 10 seconds. If
$\Gamma$
fluctuates by a factor of $\sim 2$ around its mean value, relative motions
within
the outflowing material will themselves (in the comoving frame) be
relativistic, and
can give rise to internal shocks. For $\Gamma \sim 10^2$, the resultant
dissipation
occurs outside the 'photosphere' and can convert a substantial fraction of the
overall outflow energy into non-thermal radiation.  This suggests a mechanism
for cosmological bursts that demands less extreme assumptions (in respect of
$\Gamma$-values, freedom from baryonic contamination, etc) than earlier
proposals.
\bsk
\ctl{ \bf 1. Introduction}
\bsk
Gamma ray bursts are clearly the 'signal' of an energetic event lasting
(typically) no more than a few seconds. The energy output in the gamma rays
themselves, if these events are at 'cosmological' distances, would be up to
$10^{51}\erg$. Each burst would then, probably, involve the collapse of a
stellar-mass object, or the coalescence of a compact binary (see Hartmann,
1993,
or \Pacz, 1994, for an overall review).  The main theoretical challenge is to
understand how the energy can be converted, with adequate efficiency, into
gamma
rays with a non-thermal spectrum

    A widely recognized problem is that if the rest mass energy of entrained
baryons exceeded even $10^{-5}$ of the
total energy, the associated opacity would trap the radiation so that it was
degraded by adiabatic expansion (and thermalized) before escape, e.g. Cavallo
and Rees, 1978, \Pacz, 1990. The energy would therefore be transformed into
kinetic energy of bulk relativistic outflow.  This problem arises if the
'event' is approximated as an instantaneous fireball, or as an outflowing wind
which is 'steady' over the entire burst duration.
In earlier papers, we have discussed how  kinetic energy can be reconverted
into
gamma rays  by relativistic shocks which form when the ejecta run into
external
matter (either ambient interstellar matter, or
a non-relativistic outflow preceding the 'event' itself). We showed that
acceptable models required bulk Lorentz factors $\Gamma$ of order 1000 --
still high, but allowing much more baryonic loading and opacity than would
be tolerable if the energy escaped directly from a `simple' fireball
before adiabatic losses had attenuated and thermalized it.

In this paper we show that the constraints are eased still further if we
adopt a less idealized  picture of the 'event' itself.  The  production of the
relativistic ejecta will be spread over a finite duration; moreover the
physical
conditions determining $\eta=E/M c^2$ (or $\eta=L/{\dot M}c^2$, the ratio of
radiation and magnetic energy to rest mass, which gives an upper limit
for $\Gamma$) will not be steady  throughout the event.  Fast (higher $\eta$)
ejecta
can catch up with slower material that was ejected earlier: kinetic energy can
then be reconverted into energetic particles (and thence into gamma-rays).
This suggests a mechanism for 'cosmological' bursts that can operate for much
lower values of $\eta$ (i.e.  higher loading factors)
than were previously believed necessary.

   In these more realistic models, dissipation happens whenever internal shocks
develop in the ejecta --  it need not await the deceleration by sweeping up
ambient external matter. Simple relativistic kinematics show, however,  that
the internal shocks do not develop until the ejecta have attained sufficiently
large distances that the resultant radiation can escape without thermalization
or adiabatic losses.

We show how  variations in the terminal speed, resulting from inhomogeneous
or time-varying conditions around the central object,  can yield efficient
production of gamma rays,  and generally a complex time-structure.   We outline
the general kinematics in \S 2. We then discuss in \S 3 the nature of the
dissipation, and the distinctive role of strong fields in
'magnetically-dominated' outflows. The relation of these ideas to gamma-ray
burst phenomenology is outlined in \S 4.
\bsk
\ctl{\bf 2. Kinematics in an Unsteady Relativistic Outflow}
\bsk
2.1 ~ An Illustrative Example
\msk
We postulate an outflow persisting (typically) for
a few seconds. But instead of assuming this to be a steady wind we suppose that
it is irregular on much shorter timescales. This  may be a 'wind' from a
high-$B$ newly formed pulsar (Usov 1992,94), or the debris (again probably
highly magnetized) flung off from a compact  binary during the complex dynamics
of its coalescence into a black hole and surrounding disc (Narayan, \etal,
1992,
Davies, \etal, 1994).

Whenever part of the ejecta  'catches up' with other material ejected earlier
at a lower Lorentz factor,  an internal shock forms, which dissipates the
relative kinetic energy.  To illustrate the basic idea, suppose that two
'blobs'
of equal rest mass, but with different Lorentz factors $\Gamma_1$ and
$\Gamma_2$
(with $\Gamma_2 >\Gamma_1 \gg 1$)
are ejected at times $t_1$ and $t_2$, where $t_2 - t_1 =  t_{var}$. In the case
of highly relativistic ejecta, the shock develops after
a distance of order $c t_{var} \Gamma_1 \Gamma_2$. For high Lorentz
factors, therefore, the shock takes a long time to develop, even if $\Gamma \gg
1$. This is, of course, because the distance that must be caught up
is (in the  'laboratory' frame) of order $c  t_{var}$, but the speeds
 all differ from  $c$ by less than $1/\Gamma_1^2$.

For example, suppose that the Lorentz factor of the outflow is, on average,
100, but varies from 50 to 200 on a timescale $t_{var}$. The velocity
differences are of order $10^{-4} c$, so the distance for the shock to develop
in
the lab frame is $10^4 c t_{var}$. The  reconversion of bulk energy
can nevertheless be very efficient: when the two blobs share their momentum,
they move with
$\Gamma_{final}=\sqrt{\Gamma_1\Gamma_2}$, so  the fraction of the energy
dissipated is
$$
\varepsilon=(\Gamma_1 +\Gamma_2 -2\sqrt{\Gamma_1\Gamma_2})/(\Gamma_1 +\Gamma_2)
                                                                  ~.\eqno(1)
$$
For the previous numerical example, the efficiency would be 20\%.
High efficiency  does not, therefore,  require an impact on matter
at rest; all that is needed is that the relative motions in the comoving
frame be relativistic  --i.e. $\Gamma_2/\Gamma_1 > 2$ (c.f. Rees, 1978,
for an application of this argument in a different context).
\bsk
2.2 ~ An Unsteady Wind
\msk
Suppose that the mean outflow (over the few seconds that a typical event
lasts) can be characterized by a steady wind with given mean values of
$L_w$ and $\eta= L_w/{\dot M} c^2$. We then envisage that the value of
$\eta$ (or $L_w$) is unsteady.

The mean properties of the wind determine the average bulk Lorentz factor
$$
\Gamma \sim \cases{ (r/r_l) ~,& for $r \siml r_s$;\cr
                     \eta   ~,& for $r \simg r_s$.\cr} \eqno(2)
$$
Here $r_l$ is the size at the base of the wind (in 'young pulsar' models,
c.f. Usov, 1992, this would equal $c P /2\pi$, where $P$ is the period).
The Lorentz factor
saturates to $\Gamma\sim\eta$ at a saturation radius $r_s /r_l \sim \eta$
where the wind energy density, in radiation or in magnetic fields,  drops below
the baryon rest mass density in
the comoving frame. For a given baryonic mass loss $\dot M$ the photospheric
radius where the wind becomes optically thin to Thomson scattering is
$$
 r_{ph}={\dot M} \kappa /(4\pi c \Gamma^2) =1.2\times 10^{12} L_{51}\eta_2^{-3}
                                                  ~\cm~, \eqno(3)
$$
where $\eta_2=(\eta/10^2)$ and $L_{51}=(L_w /10^{51}\ergs)$.
The above equation holds provided that $\eta$ is low enough that the wind has
already reached its 'terminal' Lorentz factor at $r_{ph}$. This requires
$\eta \siml \eta_m \sim 10^2 L_{51}^{1/4} t_{var}^{-1/4}$, if one takes
$r_l \sim ct_{var}$.

When a compact binary coalesces,  the ejecta would obviously be very messy.
The characteristic timescale at the 'base' of the wind is of order $t_{dyn}=
10^{-3}\s$. (The breakup angular speed of a pulsar is of the same order).
Large-amplitude variations could occur on any timescale longer than this. (Even
more rapid  variations are also possible, especially if there is a strong
tangled magnetic field in the wind, e.g. \S 3).

If the value of $\eta$ at the base increases by a factor $\simg 2$ over
a timescale $t_{var}$, then the later ejecta will catch up and dissipate
a significant fraction of their energy at some radius $r_d > r_s$ given by
$$
r_d \sim c t_{var} \eta^2 \sim 3\times 10^{14} t_{var} \eta_2^2 ~\cm~,
\eqno(4)
$$
where $t_{var}$ is in seconds.
Dissipation, to be  most effective, must occur when the wind is optically thin
-- otherwise it will suffer adiabatic cooling before escaping, and be
thermalized. (Outside $r_s$ and $r_{ph}$, where radiation has decoupled
from the plasma, the sound speed will be far below $c$; this also guarantees
that the relativistic internal motions in the comoving frame will lead to
shocks in the gas). This implies the following lower limit on $\eta$:
$$
\eta \simg 3\times 10^1 L_{51}^{1/5} t_{var}^{-1/5}~.\eqno(5)
$$
This simple estimate doesn't take into account the extra pairs that may
result from dissipation. We comment further on this in \S 3.

The physical conditions in these shocks qualitatively resemble those in the
reverse shocks behind the blast waves discussed by  \Mesz and Rees (1993);
however the densities and magnetic fields are higher, and the coling more
efficient, because  $r$ is smaller.  (The internal Lorentz factors
are modest, because, to first approximation, all the material is comoving
outwards at more or less the same speed).

There will be variations with a range of $t_{var}$. Rapid fluctuations are
dissipated at smaller $r$ than those on longer timescales. There would
therefore
be a dependence of the spectrum on the characteristic variability timescale .
The radiation processes depend on the magnetic field strength. We discuss
this next, because the field may also be dynamically important.
\bsk
\ctl{\bf 3.~ The Role of the Magnetic Field.}
\bsk
Ultra-intense magnetic fields are expected either in 'young pulsar' models
(Usov
1992) or if the field builds up towards equipartition by differential motions
(Narayan et al 1992) or a convective dynamo (Duncan and Thompson, 1992).
Collapsing or coalescing neutron stars  may generate fields  as high as
$B_i\sim 10^{16} G$.
Magnetic stresses could, indeed,  be dynamically dominant over the radiation:
the ratio of magnetic energy to rest-mass energy at the base of the wind
($r = r_l$) would then determine the effective value of $\eta$.  Even  a
magnetic field that was  not  dynamically-dominant would still be important
in ensuring effective cooling by cyclotron and synchrotron emission.

If the Poynting flux provides a fraction $\alpha$ of the total luminosity $L$
at the base of the wind (at $r_l\sim c t_{var}$) the magnetic field there is
$ B_l \sim 10^{10}\alpha^{1/2}L_{51}^{1/2}t_{var}^{-1} \G$.
The comoving magnetic field at the dissipation radius (4) (which always is
outside $r_s$) is
$$
B_d=B_l(r_l/r_s)^2 (r_s/r_d) \sim 10^4\alpha^{1/2}L_{51}^{1/2}t_{var}^{-1}
                                                     \eta_2^{-3} \G~.\eqno(6)
$$
If the electrons are accelerated in the dissipation shocks to a Lorentz
factor $\gamma=10^3 \gamma_3$ the ratio of the synchrotron cooling time to
the dynamic expansion time in the comoving frame is
$$
(t_{sy}/t_{ex})_d \sim 5\times 10^{-3}\alpha^{-1}L_{51}^{-1}\gamma_3^{-1}
                                t_{var} \eta_2^5 ~,\eqno(7)
$$
so a very high radiative efficiency is ensured even for $t_{var}$ as
high as seconds.

If the shock dissipation leads to photons whose energy in the comoving frame
exceeds 1 Mev, then there is the possibility of extra pair production from
photon collisions; this would modify eq.(5). The pairs  could increase the
effective photospheric radius by a factor $x = (m_p/m_e)\times$(radiative
efficiency)$\times$(fraction of radiation going into photons above 1 Mev).
For the dissipation to occur outside any possible pair-dominated  photosphere
(a requirement that may actually be unnecessary if the pairs annihilate on a
shock cooling timescale) $\eta$ would need to be higher than in eq.(5) by
a factor $x^{1/5}$. So even if $x$ were (say) 100, the minimum required $\eta$
would go up by  by less than a factor of 3.

It is clear from eq.(7) that a  magnetic field can ensure efficient cooling
even if it is not strong enough to be dynamically significant (i.e. even for
$\alpha \ll 1$). If, however, the field is dynamically significant in the wind,
then its stresses will certainly dominate the (pre-shock) gas pressure. Indeed,
in a wind with $\alpha = 1$ the magnetosonic and Alfv\'en speeds may remain
marginally relativistic even beyond $r_s$ if the field becomes predominantly
transverse. In this extreme case, magnetic fields could inhibit shock formation
unless $\eta$ varied by much more than a factor of 2. On the other hand, the
presence of a dynamically-significant and non-uniform field could actually
drive internal motions leading to dissipation even in a constant-$\eta$ wind.
Except in the special case of an aligned dipole, the magnetic field
would have reversals on a scale of order $r_l \sim cP/2\pi =5\times 10^6 P_{-3}
\cm$.  If the field inside $r_l$ has a complex (non-dipole) structure, the
reversal could be even smaller.   (Thompson, 1994,  has discussed
a detailed model where the resultant Alfven waves are dissipated via Compton
drag  when the scattering optical depth is still large.)
\bsk
\ctl{\bf 4.~ Phenomenology and Discussion}
\bsk
An unsteady (and probably magnetized) wind or fireball has the advantage that
it
can accomodate a larger amount of baryon contamination than in previous models,
while still producing a nonthermal gamma-ray burst via synchro-Compton
radiation (as in the reverse shock of \Mesz, Laguna and Rees, 1993;
c.f. also Katz, 1994) from electrons accelerated in the shock dissipation
region beyond the photosphere.

If the energy were released as an 'instantaneous' fireball, or as a steady
wind, and transformed into kinetic energy by adiabatic expansion, then
efficient reconversion into gamma rays occurs when (but not until ) enough
external matter has been swept up to decelerate it. In earlier papers, we have
shown how the resultant shocks could give rise to gamma-ray bursts  provided
that the Lorentz factors are of order 1000; complex time-structure must then
be
put down to irregularities in the ambient medium.

In the present paper, we have assumed  (undoubtedly  more realistically)
that the energy release is complex and irregular,   and shown that this
assumption admits the extra possibility that 'internal shocks' can dissipate a
substantial fraction of the kinetic energy  before the ejecta encounter the
ambient medium.  The compact object triggering the burst is likely to have a
characteristic dynamical timescale of only about $t_{dyn}\sim 10^{-3}\s$ (of
order $t_{var}\sim r_l/c$ in \S 2). But  the energy release is likely to be
more prolonged,
determined by, e.g., magnetic spindown, disk viscosity or neutrino diffusion
timescales, depending on the model.  The energy flux in the outflow, or the
value of $\eta$, could then fluctuate on all timescales from $t_{dyn}$ up to
the overall duration of the energy release, which could be many seconds.
(Indeed, some models -- e.g. rapidly spinning pulsars with complex non-dipole
fields --  permit irregularities on timescales even shorter than $t_{dyn}$).

In comparision with shocks involving external ambient matter, the internal
shocks form at smaller radii, and in regions of higher density. The  Lorentz
factors (and values of $\eta$) needed in order to get efficient dissipation,
and
short 'observer frame' timescales,  can then be somewhat lower. However, values
of $\eta \simg 30 (L_{51}/t_{var})^{1/5}$ (eq.[5]) are still required in order
to
ensure that the shocks form outside the photosphere.

This implies wind mass losses $\dot M \siml 3\times 10^{-2}
L c^{-2} \sim 10^{-5} L_{51} \msun\si$. The total (isotropic) mass loss
expected from an unmagnetized collapsing core or merging compact binary is of
order $10^{-3}\msun \si$ (e.g. \Mesz and Rees, 1992, Woosley, 1993). However,
along the rotational (or binary) axis centrifugal forces may significantly
reduce the baryon losses, while in a strongly magnetized object mass loss is
expected only from the open polar field lines, representing a fraction $\siml
10^{-2}$ of the total area, so $\dot M \siml 10^{-5}\msun$ is reasonable.
(The value of $\dot M$ implied by our required value of $\eta $ ($\sim 10^2$)
are high enough to ensure that the MHD approximation applies to any magnetic
field in the wind; this contrasts with Usov's (1992, 1994) proposal, which
would require a very much lower $\dot M$).

The  observed bursts are remarkable for their disparate and complex
time-structure, and this may be a consequence of how the development of
internal shocks is subject to irregularities in the outflow.

In summary, the implications of this work are \lbr
(i) The short timescales (and adequate efficiencies) do not need such high
values of $\eta$ as our earlier blast wave models (which were themselves much
less demanding in this respect than pre-1992 'fireball' models). \lbr
(ii) The time structure could be complex, being dependent on the time history
of
the lorentz factor. The dissipation associated with shorter timescales would
tend to occur  at smaller radii. \lbr
(iii) Since the observed gamma rays from each part of the wind are
concentrated,
owing to aberration, into an angle of order 1/$\eta$, our discussion can be
straightforwardly extended to a 'beamed' or jet-like geometry. The broad range
of burst morphologies could then , at least in part, be due to the axis and the
boundaries of jets having different Lorentz factors or internal
variability.\lbr
(iv) The energy sources in a cosmological context could be either
stellar collapse (Usov, 1992, Woosley, 1993) or compact binary mergers
(\Pacz, 1986, Eichler, \etal, 1989, Narayan, \Pacz and Piran, 1992).\lbr
%(v) The gamma rays may come from a region where  the field is high enough to
%yield in some cases cyclotron features observable in the hard X-ray band.\lbr
%
\bsk
{\it Acknowledgements}: This research has been partially supported through NASA
NAGW-1522, NAG5-2362 and by the Royal Society.
\bsk
%\newpage
\ctl{\bf References}
\bsk
\ref Cavallo, G. and Rees, M.J., 1978, M.N.R.A.S., 183, 359
\ref Davies, M.B., Benz, W., Piran, T., and Thielemann, F.K., 1994,  Ap. J., in
press.
\ref Duncan, R.C. and Thompson, C., Ap.J.(Letters), 392, L9
\ref Eichler, D., Livio, M., Piran, T. and Schramm, D., 1989, Nature, 340, 126
\ref Hartmann, D., \etal, 1993, to appear in {\it High Energy Astrophysics},
  J. Matthews, ed. (World Scientific).
\ref Katz, J.I., 1994, Ap.J., 422, 248
\ref \Mesz, P. and Rees, M.J., 1992, Ap.J., 397, 570
\ref \Mesz, P. and Rees, M.J., 1993, Ap.J., 405, 278
\ref \Mesz, P., Laguna, P. and Rees, M.J., 1993, Ap.J., 415, 181.
\ref Narayan, R., Paczynski, B. and Piran, T., 1992, Ap.J.(Letters), 395, L83
\ref Paczy\'nski, B., 1986, Ap.J.(Lett.), 308, L43
\ref Paczy\'nski, B., 1990, Ap.J., 363, 218.
\ref Paczy\'nski, B., 1994, in {\it Proc. Huntsville Gamma-ray Burst Wkshp.},
 eds. G. Fishman, K. Hurley, J. Brainerd, (AIP: New York), in press
\ref Rees, M.J., 1978, M.N.R.A.S., 184, 61P
\ref Thompson, C., 1994, M.N.R.A.S., in press
\ref Usov, V.V., 1992, Nature, 357, 472
\ref Usov, V.V., 1994, M.N.R.A.S., 267, 1035
\ref Woosley, S., 1993, Ap.J., 405, 273
\bsk
%\newpage
\ctl{\bf Figure Caption}
\bsk
Fig. 1:~~  Wind regimes as a function of $\eta=L/{\dot M}c^2$ and
$r/c t_{dyn}$. The lines above which the wind becomes optically thin to
scattering (due to $\dot M$) has a slope -1/3  or -1/2 for $L$ or $\dot M$
constant after saturation, and is fixed before saturation. The lines of
$t_{var}$ proportional to $t_{dyn}$ have a slope 1/2.
GRBs from self-consistent unsteady winds are in the
triangular region lying below the line $t_{var}\sim t_{dyn}$, and
above the line $\tau=1$ representing the photosphere.
\end